\documentclass[letter,aps,prd,10pt,preprintnumbers,showpacs,showkeys,superscriptaddress,nofootinbib,amsmath,amssymb,floatfix
]{revtex4-1}
\usepackage{times}
\usepackage{hyperref}
\usepackage{amsmath}
\usepackage{amssymb}
\usepackage{graphicx}
\usepackage{subfigure}
\usepackage{booktabs}
\usepackage{rotating}
\usepackage{longtable}
\usepackage{bm}
\usepackage{float}
\usepackage{epstopdf}
\usepackage{xcolor}
\usepackage{comment}

\makeatletter

\newcommand{\Rmnum}[1]{\expandafter\@slowromancap\romannumeral #1@}
\makeatother

\makeatletter
\newcommand*{\rom}[1]{\expandafter\@slowromancap\romannumeral #1@}
\makeatother
\begin{document}
\title{ Quasinormal Modes and Topological Characteristics of a Schwarzschild Black Hole Surrounded by the Dehnen Type Dark Matter Halo}

\author{Farokhnaz Hosseinifar}
\email{f.hoseinifar94@gmail.com}
\affiliation{Center for Theoretical Physics, Khazar University, 41 Mehseti Str., Baku, AZ1096, Azerbaijan}
\author{Shahin Mamedov}
\email{ctp@khazar.org (Corresponding author)}
\affiliation{Center for Theoretical Physics, Khazar University, 41 Mehseti Str., Baku, AZ1096, Azerbaijan}
\affiliation{ Institute for Physical Problems, Baku State University, Z.Khalilov 23, Baku, AZ 1148, Azerbaijan}
\affiliation{Institute of Physics, Ministry of Science and Education, H.Javid 33, Baku, AZ 1143, Azerbaijan}
\affiliation{Lab. for Theor. Cosmology, International Centre of Gravity and Cosmos, Tomsk State \\ University
of Control Systems and Radio Electronics (TUSUR), 634050 Tomsk, Russia}
\author{Filip Studni{\v{c}}ka}
\email{filip.studnicka@uhk.cz}
\affiliation{Department   of   Physics, Faculty of Science,  University   of   Hradec   Kr\'{a}lov\'{e}, Rokitansk\'{e}ho   62, 500   03   Hradec   Kr\'{a}lov\'{e},   Czechia}
\author{Hassan  Hassanabadi}
\email{hha1349@gmail.com}
\affiliation{Department   of   Physics, Faculty of Science,   University   of   Hradec   Kr\'{a}lov\'{e},  Rokitansk\'{e}ho 62, 500   03   Hradec   Kr\'{a}lov\'{e},   Czechia}
\affiliation{Departamento de F\'{\i}sica Te\'orica, At\'omica y Optica and Laboratory for Disruptive \\ Interdisciplinary Science (LaDIS), Universidad de Valladolid, 47011 Valladolid, Spain}

\begin{abstract}

In this work, we explore the critical parameters that delineate the existence of black holes, identifying the permissible ranges that facilitate their formation. A comprehensive thermodynamic analysis of black holes is conducted, leading to the calculation of black hole remnants. We investigate the trajectory of light, establishing an upper limit for the parameters based on Event Horizon Telescope (EHT) observations of Sgr A*, ensuring that the black hole's shadow resides within the allowed region. Furthermore, we derive the quasinormal modes (QNMs) for both scalar and electromagnetic perturbations. Utilizing a topological framework, we examine the stability of the photon sphere and classify the topology of the black hole in accordance with its thermodynamic potentials.
\end{abstract}
\maketitle
\section{Introduction}\label{Sec1}
Black holes, as the ultimate manifestation of gravitational collapse, are fascinating subjects to enhance our understanding of gravity and spacetime \cite{bronnikov2013black,malafarina2017classical}. Exploration of the environments surrounding black holes is of supreme importance in modern astrophysics \cite{volonteri2012black,alexander2012drives,volonteri2021origins}. The materials and phenomena in their vicinity, ranging from accretion disks to relativistic jets, offer critical insights into the fundamental processes of matter under extreme conditions \cite{rieger2011nonthermal,alexander2017stellar,blandford2019relativistic}. Understanding these interactions not only enhances our knowledge of black hole physics but also sheds light on the dynamics of galaxies and the universe.

Dark matter is an integral component of the universe and plays a crucial role in shaping the cosmic landscape \cite{frieman2008dark}. The gravitational effects of dark matter can significantly alter the environments around these objects, affecting their formation, growth, and observable characteristics \cite{kauffmann1993formation,capozziello2012dark,bertone2018history,konoplya2022solutions,capozziello2023dark,capozziello2023modified,shen2024analytical,molla2025observable,abbasi2011search}. There are various dark matter density profiles which describe different dark matter distributions based on model parameters \cite{ascasibar2004physical,merritt2006empirical,catena2010novel,salucci2010dark}.
The Dehnen--type dark matter halo model is particularly relevant in this context, as it provides a flexible framework for analyzing the density profiles of dark matter in various cosmological scenarios \cite{al2024quasinormal,jha2025shadow,gohain2024thermodynamics}.
The Dehnen--type halo refers to a family of spherical mass models that describes the distribution of dark matter in haloes \cite{dehnen1993family}.
Using this model, it is investigated how variations in core density and radial distribution affect the gravitational potential surrounding black holes \cite{bronnikov2007regular,khonji2024core}. 
The gravitational dynamics of a Schwarzschild-like black hole within a Dehnen-type dark matter halo, focusing on periodic orbits and gravitational wave signatures \cite{alloqulov2025gravitational}, the affect of density and radius of the dark matter halo on the black hole's shadow and photon ring using observational data \cite{luo2025shadows}, and the thermodynamic properties and optical appearance of the black hole, highlighting the influence of the dark matter halo on emission intensity and image structure \cite{rani2025thermodynamic} of such black holes are also investigated.

QNM is a characteristic frequency at which the scalar field oscillations and the rate at which it is damped \cite{konoplya2006stability,nollert1999quantifying,berti2003highly}. The QNMs provide an important insight into the black hole parameters on the effective potential \cite{ferrari1984new,yang2012quasinormal,jaramillo2021pseudospectrum}. QNMs determine by solving a Schwarzschild-like equation while satisfying certain boundary conditions \cite{konoplya2022nonoscillatory,heidari2024exploring}. Of course obtaining an exact form of solution for such Schrodinger equation is impossible and therefore, numerical and semi--analytical methods are useful to find the QNMs frequencies \cite{panotopoulos2019quasinormal,dutta2020revisiting,lagos2020anomalous}. The QNMs of black holes enveloped by this type of halo have been thoroughly investigated  \cite{al2024quasinormal,jha2025shadow,liang2025quasinormal,al2025astrophysical}.
Also in Ref. \cite{hamil2025geodesics}, the authors investigate how a Dehnen--type dark matter halo influences the QNMs of Schwarzschild black holes, revealing significant effects on their oscillation frequencies and damping rates.

This study aims to investigate the implications of Dehnen-type halo dark matter on some black hole properties. By doing so, we hope to contribute valuable insights into the dark matter's profound effects on astrophysical phenomena. 

In this study, we begin by exploring the thermodynamic properties of a black hole, which will be introduced in Sec. \ref{Sec2}. Next in Sec. \ref{Sec4}, we investigate the light trajectory and shadow of the black hole. Sec. \ref{Sec6} is dedicated to examining the QNMs using an approximative method. In Sec. \ref{Sec7}, we analyze the topological charge of the photon sphere and the thermodynamic potentials associated with this black hole. Finally, we conclude our findings in Sec. \ref{Sec12}. In this work, we adopt
natural units, in which $G=c=\hbar=1$ and signature $(- + + +)$ for the metric.
\section{Metric and Thermodynamics properties}\label{Sec2}
A Schwarzschild–like black hole spacetime surrounded by a dark matter halo that is known as Dehnen–type density profile satisfies the Einstein’s field equations. The general form of the density profile in order to provide the Dehnen--Type DM halo is defined by \cite{uktamov2025static}
\begin{equation}\label{eq1}
\rho(r)=\rho_s\left(\frac{r}{r_s}\right)^{-\gamma}\left(\left(\frac{r}{r_s}\right)^\alpha+1\right)^{\frac{\gamma-\beta}{\alpha}},
\end{equation}
that $r_s$ and $\rho_s$ indicate the central halo radius and density, respectively. Parameters $\alpha$, $\beta$, and $\gamma$ are specific parameters of the density profile and take different values for different types of haloes. As an example, for the  Navarro-Frenk-White halo model which is defined by a density function with with a specific scale radius and a steep central cusp, the set $(1,\,3,\,1)$ is selected \cite{navarro1996structure}; for the Moore profile, the set $(1.5,\,3,\,1.5)$ is chosen \cite{moore1999cold}; and for the Kravtsov halo that has a flat core, the set $(2,\,3,\,0.4)$ is utilized \cite{kravtsov1998cores}.

\begin{figure}[ht!]
  \includegraphics[width=8.5cm]{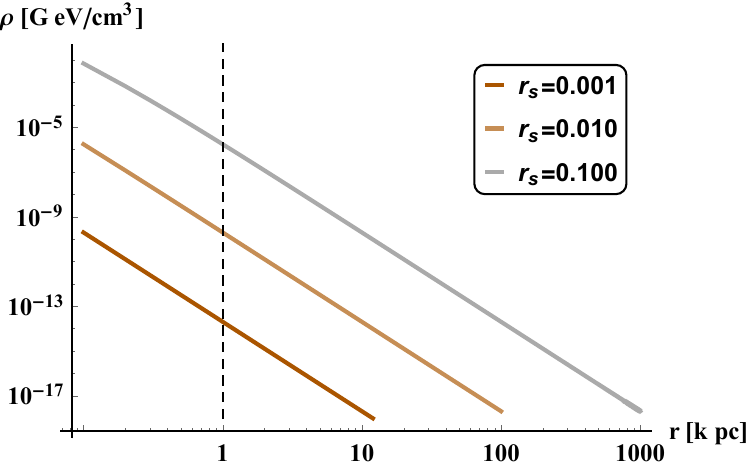} \hspace{-0.2cm}\\
    \caption{Density profile in terms of $r$ considering $\rho_s=0.02$.}\label{fig:density}
\end{figure}

Density profile of the Dehnen type dark matter halo \cite{mo2010galaxy} for the case Denhen-$(\alpha,\beta,\gamma)=(1,4,5/2)$ is considered as \cite{al2024schwarzschild}
\begin{eqnarray}\label{den}
\rho=\dfrac{\rho_s}{\left(\frac{r}{r_s}\right)^{5/2}\left(\frac{r}{r_s}+1\right)^{3/2}}.
\end{eqnarray}
In fact, by changing the values of $\alpha$, $\beta$, and $\gamma$ we change the flat core profile to cusp profile or even to the super--cusp profile. In this case, the steep inner slope is $\rho\propto r^{-5/2}$ and the outer slope is $\rho\propto r^{-4}$ that it is steeper than the previous selection of $\gamma$.
Fig. \ref{fig:density} displays the density profile
for a specific set of initial values,indicating that as $r$ approaches infinity, it tends toward zero. Also,  the effects of variation in $r_s$ on the density can be seen.
Specifically, at $r=1 k pc$ and for $\rho_s=0.02 G\,eV/cm^3$, we observe that the density is significantly dependent on changes of $r_s$. For $r_s=0.001 k pc$, the density is $\rho=1.997\times 10^{-14} G\,eV/cm^3$;  for $r_s=0.010 k pc$, the density is $\rho=1.970\times 10^{-10} G\,eV/cm^3$; and for $r_s=0.100 k pc$ it reaches $\rho=1.734\times 10^{-6} G\,eV/cm^3$.
\\Mass profile using Eq. \eqref{den} is obtained as \cite{al2024schwarzschild}
\begin{eqnarray}
M_D=4\pi\int_{0}^{r}\dfrac{\rho_s r^{'2}}{\left(\frac{r^{'}}{r_s}\right)^{5/2}\left(\frac{r^{'}}{r_s}+1\right)^{3/2}} dr^{'}=\frac{8\,\pi\,\rho_s\,r_s}{\sqrt{1+\dfrac{r_s}{r}}}.
\end{eqnarray}
The tangential velocity of the particle moving in the dark matter halo, for a spherically symmetric spacetime of form
\begin{align}\label{ds2}
ds^2=-f(r)dt^2+\frac{1}{f(r)}dr^2+h(r)\left( d \theta^2+ \sin ^2 \theta d \phi^2\right),
\end{align}
where \cite{al2024schwarzschild}
\begin{align}
f(r)=& 1-\frac{2 M}{r}-32\,\pi\,\rho_s\,r_s^2\,\sqrt{\frac{r+r_s}{r}}\label{fr},\\
h(r) =& r^2\label{hr}.
\end{align}
$f(r)$ has one root obtained from $f(r_h)=0$, in which $r_h$ refers to the horizon radius \cite{al2024schwarzschild}
\begin{equation}\label{rh}
r_h=\frac{2\,M+512\,\pi^2\,\rho_s^2\,r_s^5+32\,\pi\,\rho_s\,r_s^2\sqrt{2\,M\,(2\,M+r_s)+256\,\pi^2\,\rho_s^2\,r_s^{6}}}{1-1024\,\pi^2\,\rho_s^2\,r_s^4}.
\end{equation}
The condition for the horizon to remain positive is $32\,\pi\,r_s^2\,\rho_s<1$.
\begin{figure}[ht!]
  \includegraphics[width=8.5cm]{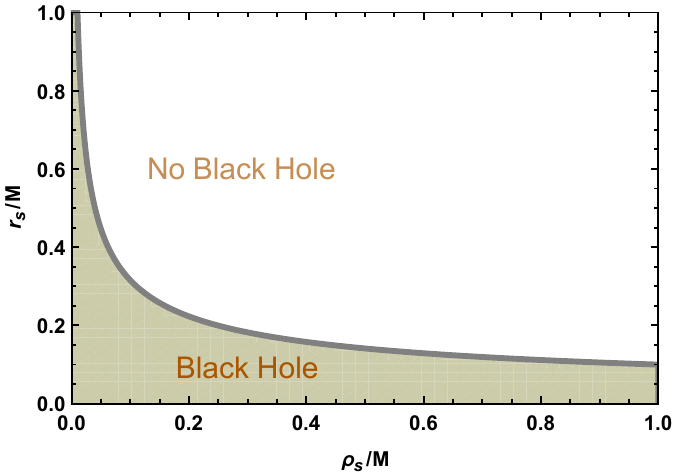} \hspace{-0.2cm}\\
    \caption{Region that black hole does exist.}\label{fig:bhnobh}
\end{figure}
Fig. \ref{fig:bhnobh} depicts the region in which the black hole exists, illustrating how this area varies with changes in the parameters $r_s$ and $\rho_s$.

Black hole's mass as a function of the horizon radius reads
\begin{eqnarray}\label{Mh}
M=\frac{r_h}{2}-16\,\pi\,\rho_s\,r_s^2\,\sqrt{r_h(r_{h}+r_s)},
\end{eqnarray}
and when $r_s$ and $\rho_s$ approach zero, the mass of the black converges to that of the Schwarzschild black hole.

\begin{figure}[ht!]
  \includegraphics[width=8.5cm]{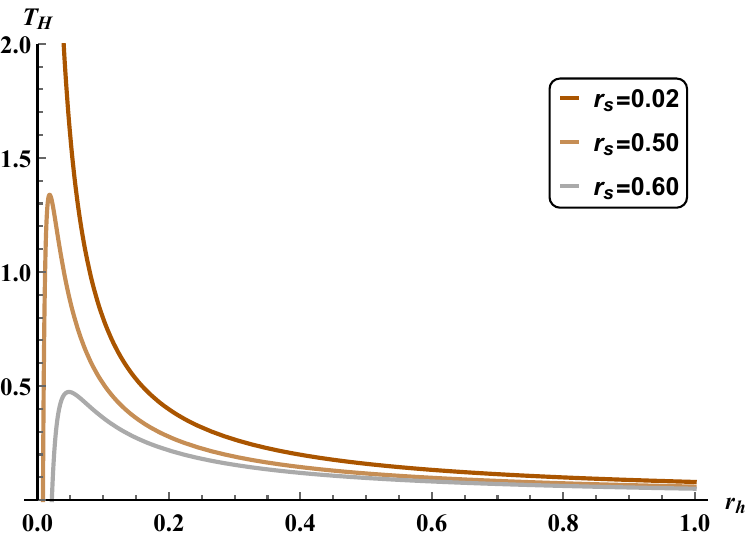} \hspace{-0.2cm}\\
    \caption{Hawking temperature as function of horizon radius for the case $\rho_s=0.01$.}\label{fig:TH}
\end{figure}

Hawking temperature for the spherically symmetric black hole of form Eq. \ref{ds2}, and by substituting the mass from Eq. \ref{Mh} is obtained as \cite{hawking1974black}
\begin{eqnarray}\label{TH}
T_H=\frac{1}{4\pi}\partial_r f(r)\bigg|_{r=r_{h}}
=\frac{1}{4\,\pi\,r_{h}}-\frac{4\,\rho_s\,r_s^2\,(2 r_{h}+r_s)}{r_{h}\sqrt{r_{h}\,(r_{h}+r_s)}}.
\end{eqnarray}
In the case where $r_s$ and $\rho_s$ tend toward zero, the temperature of the black hole approaches Hawking temperature of the Schwarzschild black hole.
Fig. \ref{fig:TH} shows the Hawking temperature curve as a function of horizon radius, revealing a second order phase transition for some choices of $r_s$ and $\rho_S$ occurring at $\partial_{r_h}T_H\big|_{r_h=r_c}=0$.
When the $\rho_s$ parameter is constant, an increase in the $r_s$ parameter results in the phase transition of the temperature occurring in a larger horizon radius.

A remnant is defined as the stable state of a black hole that remains after it has evaporated due to Hawking radiation. This phenomenon occurs during the evaporation process of a black hole which causes it to lose mass and energy over time \cite{delhom2019absorption,ong2024case}.

Remnant radius of the black hole is calculated from $T_H\big|_{r=r_{rem}}=0$ \cite{araujo2023thermodynamics} and reads 
\begin{align}
r_{rem}=\frac{r_s}{2} \left(\frac{1}{\sqrt{1-1024 \pi ^2 \rho_s^2 r_s^4}}-1\right)
\end{align}
which indicates that the condition for the existence of the remnant radius is that $32\,\pi\,r_s^2\,\rho_s<1$.

The entropy of the black hole is a measure of the uncertainty of its internal states \cite{davies1978thermodynamics,tsallis2019black}, and in accordance with the first low of thermodynamics and using Eqs. \eqref{Mh} and \eqref{TH}, is obtained as \cite{thomas2012thermodynamics,de2012black}
\begin{align}\label{entr}
S=\int\frac{\text{d} M}{T_H}=\pi\,r_{h}^2.
\end{align}
Although the relationship between entropy and horizon radius is similar to that for the Schwarzschild black hole, $r_h$ in Eq. \eqref{entr}, as observed in Eq. \eqref{rh}, depends on parameters $r_s$ and $\rho_s$.
\section{Shadow Radius}\label{Sec4}
The motion of light rays around the black hole can be investigated using the Lagrangian approach \cite{compere2007symmetries,frolov2003particle} and  in the spacetime Eq. \eqref{ds2}, the Lagrangian for the equation of motion  will take a form
\begin{eqnarray}
\mathcal{L}(x,\,\dot{x})=\frac{1}{2}(g_{\mu\nu}\dot{x}^\mu \dot{x}^{\nu})=\frac{1}{2}(-f(r)\dot{t}^2+\frac{\dot{r}^2}{f(r)}+h(r)\dot{\theta}^2+h(r)\sin^2\theta \dot{\phi}^2). \label{Lagrangian}
\end{eqnarray}
and the dot on top denotes $d/d\tau$, where $\tau$ is the affine
parameter. Using the Euler-Lagrange method, which is written as \cite{he2016gravitational}
\begin{align}\label{dL}
\frac{d}{d \tau}\left(\frac{\partial \mathcal{L}}{\partial\dot{x}^\mu}\right)-\frac{\partial \mathcal{L}}{\partial x^\mu}=0,
\end{align}
and at $\theta = \pi/2$, defining two constants of motion as $f(r)\dot{t}=E$ and $h(r)\dot{\phi}=L$, where $E$ represents energy and $L$ refers to angular momentum, 
and by substituting the conserved quantities $E$ and $L$, of eq. \eqref{Lagrangian} into Eq. \ref{dL} , we obtain the orbit equation for photons as

\begin{align}
\dot{r}^2=E^2-\frac{L^2}{h(r)}f(r)\label{Ve}.
\end{align}
We obtain the following equation for the trajectory of light rays
\begin{align}
\left(\frac{dr}{d\phi}\right)^2&=h(r) f(r)\left(\frac{h(r)}{f(r)}\frac{E^2}{L^2}-1\right).\label{traj}
\end{align}
The term $\left(dr/d\phi\right)^2$ is considered as the effective potential and due to the photon circular orbit,
\begin{equation}\label{VdV}
V_{\rm eff}\big|_{r=r_{ph}}=0\;\text{and}\;\partial_r V_{\rm eff}\big|_{r=r_{ph}}=0,
\end{equation}
where $r_{ph}$ indicates the photon radius of the black hole, and the term $L/b$ can be calculated as
\begin{equation}\label{b}
\frac{L^2}{E^2}=\frac{h(r_{ph})}{f(r_{ph})}=\frac{r_{ph}^3}{r_{ph}-2\,M-32\,\pi\,\rho_s\,r_s^2 \sqrt{r_{ph}(r_{ph}+r_s)}}
.
\end{equation}
Also, the photon sphere of a black hole is a spherical region around the black hole, where gravity is strong enough that photons can move in orbits \cite{gralla2020shape,perlick2022calculating}. The photon sphere radius due to the circular orbit of light rays using Eq. \eqref{VdV} is computed from the relation \cite{virbhadra2000schwarzschild,claudel2001geometry}
\begin{equation}
\left(r^2\,\partial_r f(r)- 2\,r\,f(r)\right)_{r=r_{ph}}=0,\label{rph}
\end{equation}
from which we obtain  explicitly
\begin{equation}\label{rph2}
6 M+16 \pi \rho_s r_s^2 (4 r_{ph}+5 r_s) \sqrt{\frac{r_{ph}}{r_{ph}+r_s}}=2 r_{ph}.
\end{equation}
An unstable photon sphere represents a critical boundary where light can either escape the black hole or be captured by its gravitational pull, which leads to the formation of observable phenomena \cite{gralla2020shape,perlick2022calculating}.

\begin{figure}[ht!]
  \includegraphics[width=8.5cm]{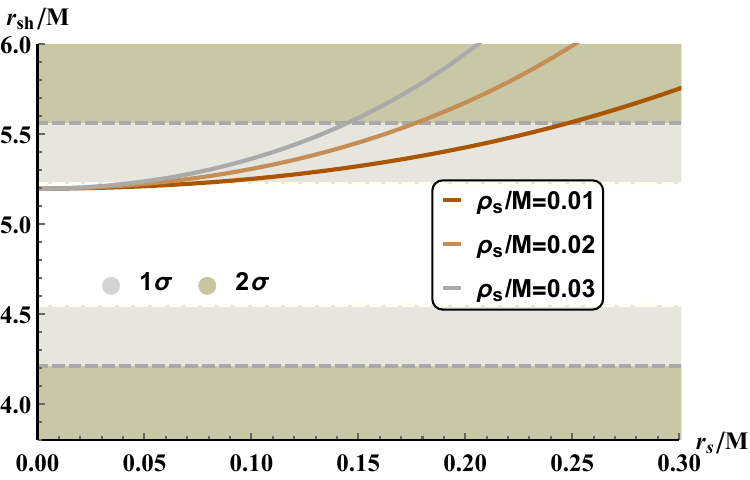} \hspace{-0.2cm}\\
    \caption{Shadow radius by varying parameter $r_s$ for three cases of parameter $\rho_s$. White region is allowed for $1\sigma$ region and both white and light and light eggshell color are allowed for $2\sigma$ region.}\label{fig:Shad}
\end{figure}

The shadow radius indicates the size of the dark region observed when the black hole is positioned between an observer and a light source. This shadow is essentially the result of gravitational lensing, where light from the source is bent around the black hole \cite{atamurotov2015optical,jusufi2020quasinormal,zakharov2012shadows}.
For an observer who is located at $r\rightarrow\infty$, shadow radius of the black hole of form Eq. \eqref{ds2} is given by \cite{perlick2022calculating,perlick2018black}
\begin{equation}
r_{sh}=\dfrac{r_{ph}}{\sqrt{f(r_{ph})}}\sqrt{1-32\,\pi\,\rho_s\,r_s^2}
=\frac{r_{ph}^{3/2} \sqrt{1-32\,\pi\,\rho_s\,r_s^2}}{\sqrt{r_{ph}-2\,M-32\,\pi\,\rho_s\,r_s^2\,\sqrt{r_{ph}\,(r_{ph}+r_s)}}}
\end{equation}
Fig. \ref{fig:Shad} illustrates shadow curves by varying parameter $r_s/M$ for three cases of parameter $\rho_s/M$.
The reported results from the EHT indicates that the shadow radius of Sgr A* for the $1\sigma$ region, should be within the $4.55<r_{sh}/M<5.22$ range, and for the $2\sigma$ region, should be within the $4.21<r_{sh}/M<5.56$ range \cite{vagnozzi2023horizon}. Thus, some constraints can be found for the black hole parameters of metric Eq. \eqref{ds2} in comparison with EHT data. Considering three cases of $\rho_s/M$, the allowable range for parameter $r_s/M$ is reported in Table \ref{Table:rshl}.
\begin{table}[ht!]
		\caption{The allowed range of parameter $r_s/M$, $\text{(upper bound, lower bound)}$, based on the obtained documentation to observations of Sgr A* BH, for different cases of $\rho_s/M$.}
		\centering
		\label{Table:rshl}
		\begin{tabular}{|c|c|c|}
		\hline
		 $\rho_s/M$ & $1\sigma$ & $2\sigma$ \\
		 \cline{2-3}
		 \hline
		 $0.01$ & $\;(0.0,\,0.07)\;$ & $\;(0.0,\,0.25)\;$\\
		 \hline
		 $0.02$ & $\;(0.0,\,0.05)\;$ & $\;(0.0,\,0.18)\;$\\
		\hline
		 $0.03$ & $\;(0.0,\,0.04)\;$ & $\;(0.0,\,0.14)\;$\\
		 \hline
		\end{tabular}
	\end{table}
As anticipated, when $r_s/M$ and $\rho_s/M$ approach zero, the radius of the black hole's shadow converges to that of the Schwarzschild black hole, reaching a value of $3\sqrt{3}M$. Consequently, as illustrated in Fig. \ref{fig:Shad}, there is no lower bound for the parameters that would place the shadow within the permissible region. Increasing $r_s/M$ while keeping $\rho_s/M$ constant results in a larger shadow radius. Additionally, it is clear that with a fixed $r_s/M$, an increase in $\rho_s/M$ also leads to an expansion of the black hole's shadow. Therefore, we can conclude that as $\rho_s/M$ increases, the virtual region of the parameter $r_s/M$ for which the shadow resides within either the $1\sigma$ or $2\sigma$ regions diminishes.


\section{Quasi-Normal Modes by WKB method}\label{Sec6}
QNMs represent the natural frequencies of oscillation of black holes and compact objects in general relativity when they are subjected to perturbations, such as gravitational waves \cite{nollert1999quasinormal,cavalcanti2022echoes}.
The significance of the QNMs lies in their unique properties; they are characterized by complex frequencies, where the real part indicates the oscillation frequency and the imaginary part represents the damping of the oscillation due to energy dissipation \cite{alpeggiani2017quasinormal,lalanne2019quasinormal,sauvan2022normalization}. The concept of QNMs was first introduced in the context of black holes by Vishveshwara \cite{vishveshwara1970scattering}, who noted their relevance to gravitational wave scattering and the stability of black holes \cite{kokkotas1999quasi,hegde2019quasinormal}. The study of QNMs encompasses perturbations of various types of black holes, including Schwarzschild, Reissner-Nordstr{\"o}m, Kerr, and Kerr-Newman black holes, as well as neutron stars \cite{blazquez2019quasinormal,cai2021quasinormal,franchini2024testing}.

To investigate the relativistic spinless massless scalar fields we apply the Klein-Gordon equation. As the QNMs are often investigated in the context of scalar perturbations near a black hole, therefore choosing the Klein-Gordon equation is a suitable choice to model these kinds of perturbations, which reads \cite{heidari2023investigation}
%
\begin{align}
\frac{1}{\sqrt{-g}}\partial_\mu(\sqrt{-g}g^{\mu\nu}\partial_\nu\Psi_{\omega lm}(\textbf{r},t))=0.
\end{align}
The QNM frequencies are determined by finding solutions to the wave equation with boundary conditions at the event horizon and spatial infinity \cite{kovtun2005quasinormal,konoplya2011quasinormal}.
By separating the radial and angular components, the wave function can be assumed as

\begin{align}\label{psi}
\Psi_{\omega lm}(\textbf{r},t)=\frac{R_{\omega l}(r)}{r}Y_{lm}(\theta,\,\phi)e^{i\omega t},
\end{align}
where $Y_{lm}(\theta,\,\phi)$ indicates the spherical harmonics and $\omega$ represents the frequency. Thus, an equation for the radial part of Eq. \eqref{psi} for a metric of the form Eq. \eqref{ds2} is obtained as
\begin{align}\label{psir}
f(r)\frac{d}{dr}(f(r)\frac{dR_{\omega l }(r)}{dr})+(\omega^2-V(r))R_{\omega l}(r)=0.
\end{align}
Eq. \eqref{psir} in the tortoise coordinates $dr^*=dr/f(r)$ reads
\begin{align}
\frac{d^2}{dr^{*^2}}R_{\omega l}(r^*)+(\omega^2-V_{\text{eff}}(r^*))R_{\omega l}(r^*)=0,
\end{align}
and the effective potential is considered as
\begin{align}
V_{\text{eff}}(r)=\frac{f(r)(1-s^2)}{r}\frac{df(r)}{dr}+\frac{f(r)l(l+1)}{r^2},
\end{align}
where $s$ have values of 0,1,1/2, and 2 for scalar, electromagnetic, fermion, and graviton fields, respectively.
\\Several methods are employed to calculate QNMs frequencies. Numerical integration techniques are commonly used to solve wave equations for specific potentials. High-precision numerical methods have shown good agreement with the analytical results derived from perturbation theory \cite{kokkotas1999quasi,volkel2022quasinormal,cho2012new}. The Wentzel-Kramers-Brillouin (WKB) method provides an approximation for calculating the frequencies of the QNMs by treating the wave equation as a potential scattering problem. This method is particularly useful for estimating frequencies near the peak of the effective potential \cite{schutz1985black,guinn1990high,jusufi2020quasinormal,konoplya2021black}. 
The continued fraction techniques involves representing the eigenvalue problem in terms of a continued fraction, which can be solved iteratively to obtain the QNMs frequencies \cite{leaver1986spectral,leaver1990quasinormal,konoplya2003quasinormal,matyjasek2017quasinormal}

\begin{figure}[ht!]
  \includegraphics[width=8.5cm]{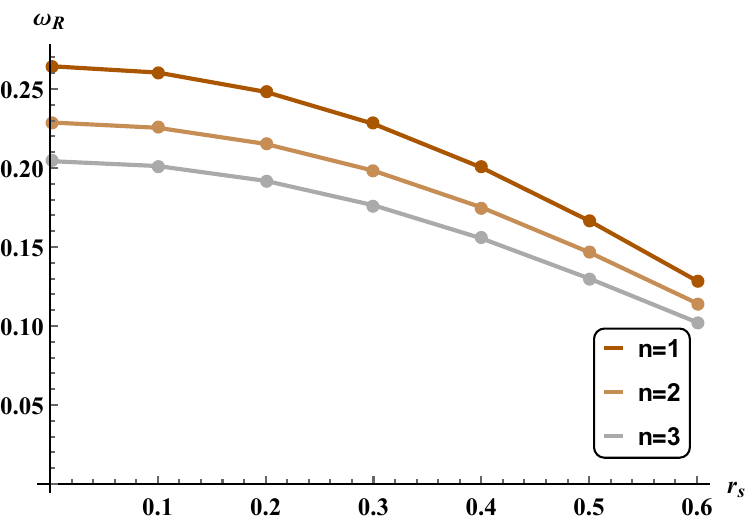} \hspace{0.5cm}
  \includegraphics[width=8.5cm]{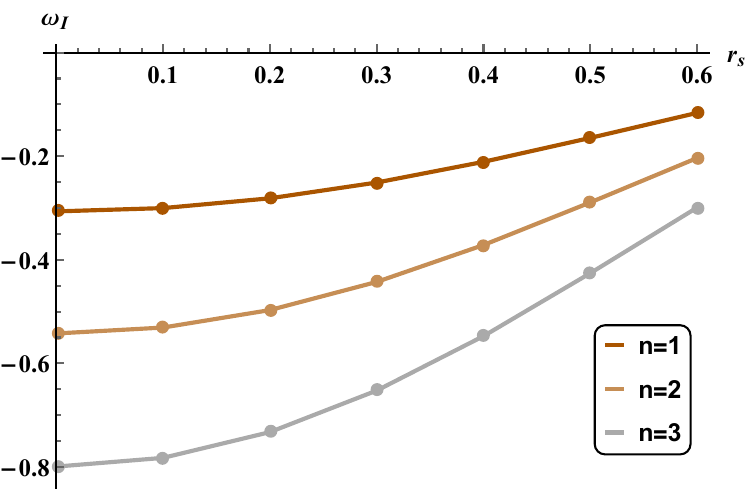} \hspace{-0.2cm}\\
    \caption{QNMs by varying parameter $r_s$, for the case $M=1$, $\rho_s=0.01$, $s=0$ and $l=1$. Left panel: real part, right panel: imaginary part.}\label{fig:OI}
\end{figure}

\begin{figure}[ht!]
  \includegraphics[width=8.5cm]{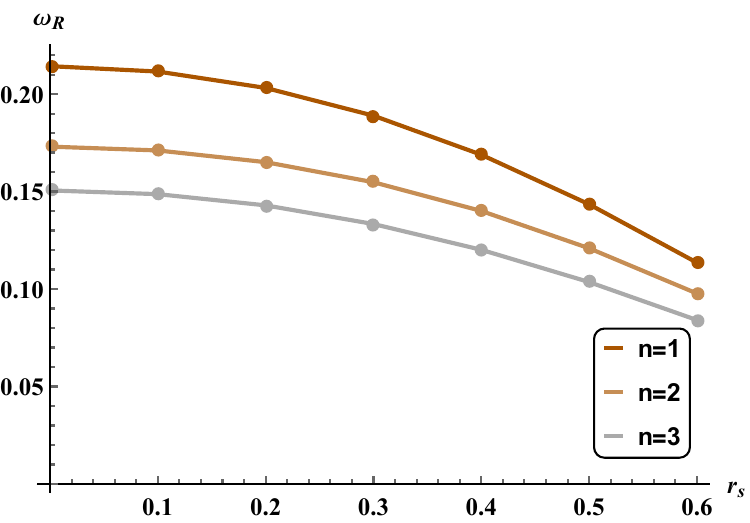} \hspace{0.5cm}
  \includegraphics[width=8.5cm]{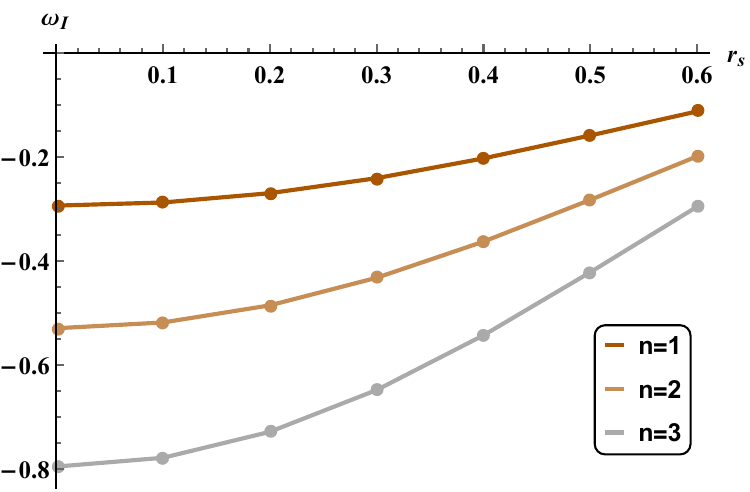} \hspace{-0.2cm}\\
    \caption{QNMs in terms of $r_s$, for electromagnetic perturbation considering $M=1$, $\rho_s=0.01$, and $l=1$. Left panel: real part, right panel: imaginary part.}\label{fig:OIs1}
\end{figure}

In this work, we have used the WKB method of
13th order and Pad\'{e} approximations introduced in Refs. \cite{konoplya2019higher,Mathematica} to calculate the QNMs of the black hole. Fig. \ref{fig:OI} shows real and imaginary parts of this oscillatory mode for the cases $M=1$, $\rho_s=0.01$, and $l=1$.
QNMs by varying $r_s$ considering $M=1$, $\rho_s=0.01$, and $l=1$ for scalar and electromagnetic perturbations are displayed in Figs. \ref{fig:OI} and \ref{fig:OIs1}, respectively.
It is evident that as $r_s$ increases, while keeping the other parameters constant, both the real part of the QNMs and the magnitude of the imaginary part decrease. This indicates that an increase in the parameter $r_s$ results in a reduction of the damping of oscillatory waves. Furthermore, it can be inferred that by maintaining the other parameters fixed, an increase in the integer $n$ leads to a decrease in the real part and an increase in the magnitude of the imaginary part, thereby increasing the damping rate in this scenario. Additionally, it is observed that as $r_s$ increases, the disparity in the magnitudes of the real and imaginary parts of the oscillatory modes for different values of $n$ diminishes. This suggests that for larger values of $r_s$, the difference in the magnitudes of the real (or imaginary) parts of the QNM across various $n$ values becomes less pronounced.
Additionally, by comparing the values of QNM in the cases of scalar and electromagnetic scattering, it can be observed that in the case of $s=1$, both the real and imaginary parts of the QNMs decrease compared to the case of $s=0$.


\section{Topology and Topological Charge of Photon Sphere and Thermodynamic Potentials}\label{Sec7}
One of the additional aspects of black hole research involves exploring them from a topological perspective \cite{s2024effective,du2024topological,wu2024novel}. 
This approach is employed to investigate the stability, the existence of phase transitions, and the topological classification of black holes. To achieve this, a potential dependent on the property under study is defined. Subsequently, vectors that represent this potential within the vector space are established. In this vector space, there may exist points referred to as zero points, where the vectors exhibit convergence or divergence. Theses points are regarded as topological defects and can be assigned a topological charge. The value of this topological charge determines the stability or topological class of the respective zero point \cite{hazarika2024thermodynamic,malik2024thermodynamic,barzi2024renyi}. This methodology is then utilized to examine the stability or instability of the photon sphere of the black hole, as well as to explore the type of phase transition at the Hawking temperature of the black hole and its topological classification based on its generalized free energy. Subsequently, we will examine the aforementioned properties from the topological perspective for the metric of Eq. \eqref{den}.
\subsection{Photon Sphere Potential Field}
\begin{figure}[ht!]
  \includegraphics[width=8.5cm]{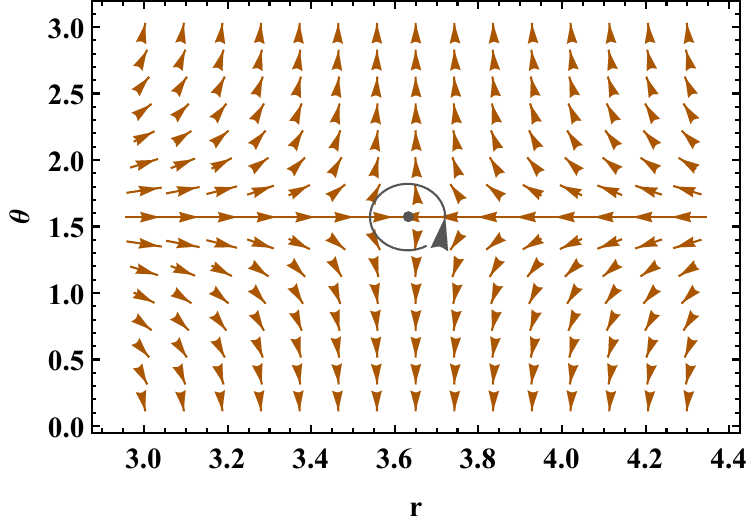} \hspace{0.5cm}
  \includegraphics[width=8.5cm]{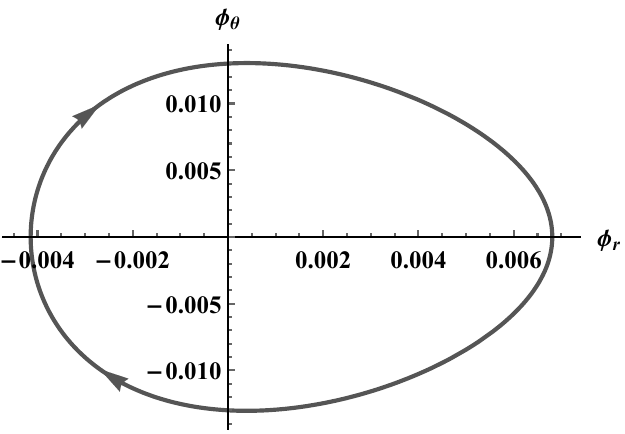} \hspace{-0.2cm}\\
    \caption{Parameters are set as $M=1$, $r_s=0.4$ and $\rho_s=0.01$. Left panel: Vector space of the potential $H(r,\,\theta)$, which includes a zero point at $r_{ph_c}=3.63057$. Right panel: $\phi_r-\phi_\theta$ curve around $r_{ps}$, which indicates a topological charge of $-1$.}\label{fig:Topomet}
\end{figure}

To study the stability of each photon sphere, a topological approach is employed, and a potential field is defined as \cite{sadeghi2024thermodynamic,wei2020topological,liu2024light, Umair}
\begin{eqnarray}\label{HPot}
H(r,\theta)=\frac{1}{\sin\theta}\left(\frac{f(r)}{h(r)}\right)^{1/2},
\end{eqnarray} 
where its relevant vectors in polar coordinates are considered as \cite{wei2020topological}
\begin{align}
\phi_{r}^H =\sqrt{f(r)} \partial_r H(r,\theta),\;
\phi_{\theta}^H =\frac{1}{\sqrt{h(r)}}\partial_{\theta} H(r,\theta).
\end{align}
As $\phi=\phi_r+i\,\phi_\theta$, the normalized form of the above vectors is obtained from \cite{wei2020topological,sadeghi2024role,zhu2024topological,wu2023classifying}
\begin{eqnarray}
n^H_r=\frac{\phi^H_{r}}{||\phi||},\;
n^H_{\theta}=\frac{\phi^H_{\theta}}{||\phi||}.
\end{eqnarray}
Zero points of the potential $H(r,\,\theta)$ are observed at $n^H_r=0$ and $\theta=\pi/2$ in the vector space, where the photon spheres of the black hole are located.
\\In Fig. \ref{fig:Topomet}, the vector field of the potential $H$ is depicted.
As shown, only one photon sphere is observed for the $r_s=0.4$ and $\rho_s=0.01$ values, which is located at the point $r_{ps}=3.63057$.
To obtain the topological charge, the $\phi_r-\phi_\theta$ curve is studied in a closed curve $c$.
If there is no zero point in the closed curve $c$, the topological charge equals zero. The counterclockwise direction of $\phi_r-\phi_\theta$ refers to a topological charge of $+1$, while the clockwise direction of the field lines corresponds to a topological charge of $-1$ \cite{rizwan2025universal}. Therefore, it can be concluded that the photon sphere located at the point $r_{ps}$ is unstable. This conclusion is consistent with what we observed in Sec. \ref{Sec4}.
\subsection{Temperature Potential Field}
\begin{figure}[ht!]
  \includegraphics[width=8.5cm]{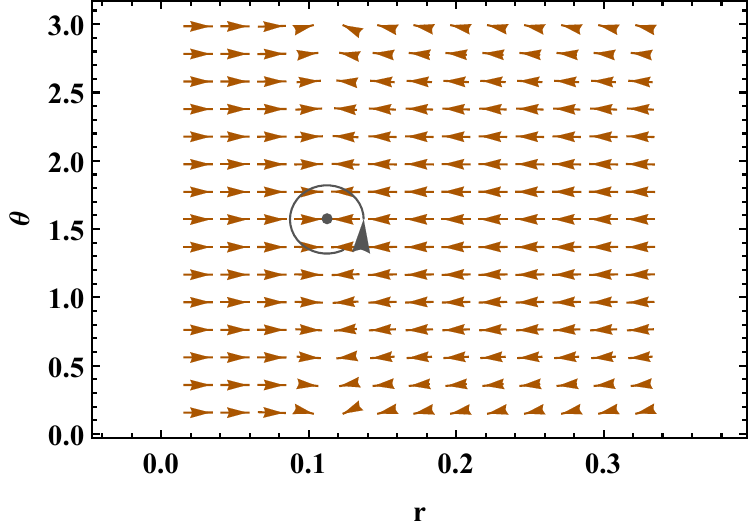} \hspace{0.5cm}
  \includegraphics[width=8.5cm]{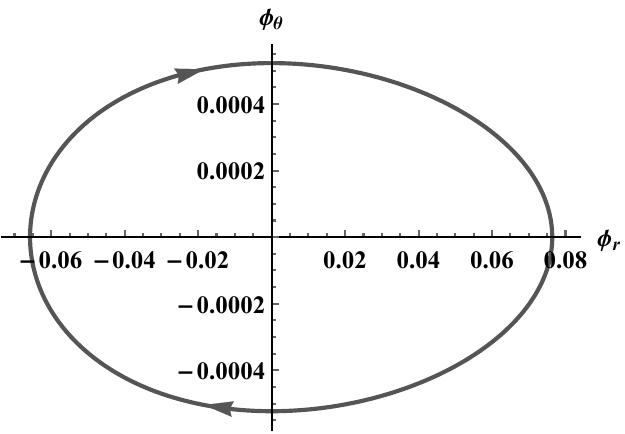} \hspace{-0.2cm}\\
    \caption{Values of $r_s=0.7$ and $\rho_s=0.01$ are given.
    Left panel: Vector space of the potential $\Phi$, which includes a zero point at $r_{h_c}=0.11229$. Right panel: Variations of $\phi_r-\phi_\theta$, that indicates a topological charge of $-1$.}\label{fig:TopoTemp}
\end{figure}

The Hawking temperature of the black hole can also be studied from a topological perspective  \cite{chen2024thermal,wei2022topology,yerra2022topology}.
In this section, similar to the previous one, a vector potential dependent on the black hole temperature is introduced. By convention, a topological charge of $-1$ refers to the conventional topological class, if the topological charge of $+1$ refers to the novel topological class.
The temperature-dependent potential is defined as \cite{wei2022topology,chen2024thermodynamic}
\begin{eqnarray}
\Phi =\frac{1}{\sin \theta} T_H,
\end{eqnarray}
and the field vectors of this potential are expressed as \cite{yerra2023topology,dong2025thermodynamic}
\begin{equation}
\phi_r^\Phi=\partial_{r_h}\Phi,\;\;\phi_{\theta}^\Phi=\partial_{\theta}\Phi.
\end{equation}
As mentioned in Sec. \ref{Sec2}, considering the $32\,\pi\,r_s^2\,\rho_s<1$ condition, there is a phase transition at the black hole Hawking temperature.
Fig. \ref{fig:TopoTemp} illustrates the normalized form of this vector field for cases $r_s=0.7$ and $\rho_s=0.01$.
In this figure, a critical point with a topological charge of $-1$ is observed at $r_{h_c}=0.11229$, which represents a conventional critical point.
\subsection{Generalized Free Energy Potential Field}
\begin{figure}[ht!]
  \includegraphics[width=8.5cm]{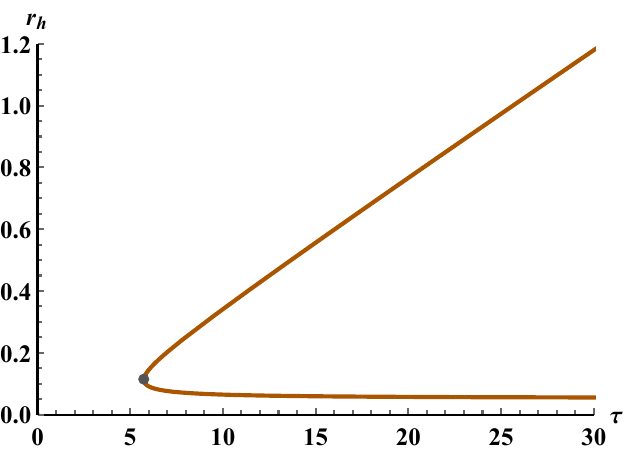} \hspace{-0.2cm}\\
    \caption{$r_h-\tau$ curve for the case $r_s=0.7$ and $\rho_s=0.01$.}\label{fig:TopoTau}
\end{figure}

\begin{figure}[ht!]
  \includegraphics[width=8.5cm]{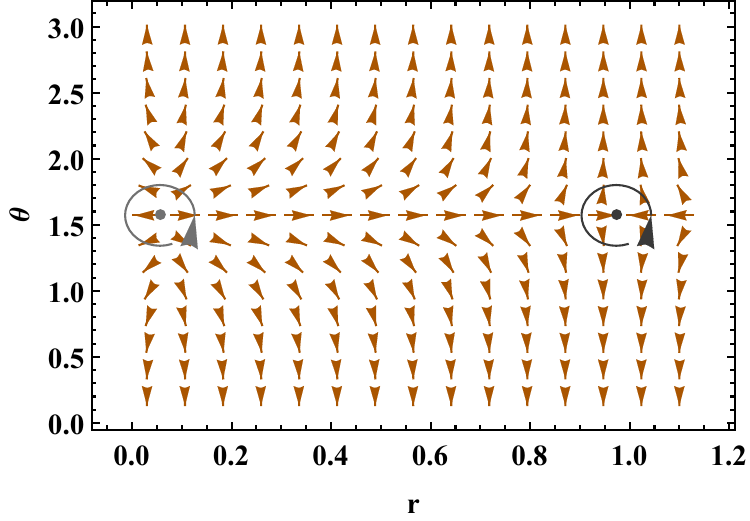} \hspace{0.5cm}
  \includegraphics[width=8.5cm]{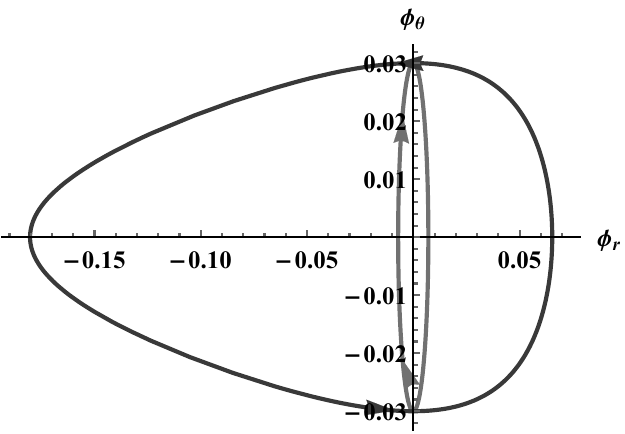} \hspace{-0.2cm}\\
    \caption{The quantities of $r_s=0.7$, $\rho_s=0.01$, and $\tau=25$ are considered. Left panel: Vector space of off-shell generalized free energy, which contains two zero points at $r_{hc_1}=0.05602$ and $r_{hc_2}=0.97324$. Right panel: The changes in $\phi_\theta$ as a function of $\phi_r$ near the zero points reveal that the topological charge at  $r_{hc_1}$ is $+1$ whereas at $r_{hc_2}$ is $-1$.}\label{fig:TopoHelm}
\end{figure}

Black hole solutions can also be considered as defects within the thermodynamic parameter space \cite{wu2024novel,wei2024universal,wei2022black,wu2023topologicall}. To this end, the generalized free energy of the black hole is defined as \cite{wei2022black,du2024topological,wu2023topological,liu2023topological}
\begin{align}
\mathcal{F}=& M_h-\frac{S} {\tau },
\end{align}
where $M_h$ refers to the mass of the black hole, and the parameter $\tau$ is considered as the inverse temperature of the cavity encompassing the black hole. The parameter $S$ denotes the entropy of the black hole, which is expressed as $\pi r_h^2$ according to the metric geometry, thus generalized free energy reads
\begin{align}
\mathcal{F}=&\frac{r_h}{2}-16\,\pi\,\rho_s\,r_h\,r_s^2 \sqrt{\frac{r_h+r_s}{r_h}}-\frac{\pi\,r_h^2}{\tau }.
\end{align}
The vectors corresponding to this potential are written as \cite{wei2022black,yasir2024topological,fan2023topological,wu2024topological}
\begin{equation}\label{dF}
\phi_r^\mathcal{F}=\partial_{r_h}\mathcal{F},\;
\phi_{\theta}^\mathcal{F}=-\cot \theta \csc \theta.
\end{equation}
To calculate the critical point of this potential, Eq. \eqref{dF} is set to zero, allowing us to express the parameter $\tau$ in terms of the horizon radius 
\begin{align}
\tau=\frac{4\,\pi\,r_h \sqrt{r_h\,(r_h+r_s)}}{\sqrt{r_h\,(r_h+r_s)-16\,\pi\,\rho_s\,r_s^2\,(2\,r_h+r_s)}}.
\end{align}
In Fig. \ref{fig:TopoTau}, the horizon radius is depicted as a function of the parameter $\tau$ for $r_s=0.7$ and $\rho_s=0.01$.
As can be seen, this curve changes direction at the point $\tau_c=5.73$. Therefore, considering $\tau>\tau_c$, the vector field of the generalized free energy potential for case $r_s=0.7$, $\rho_s=0.01$ and $\tau=25$ is illustrated in Fig. \ref{fig:TopoHelm}.
As shown, this field has two zero points located at $r_{hc_1}=0.05602$ and $r_{hc_2}=0.97324$, whose topological charges are $+1$ and $-1$, respectively. Therefore, it can be stated that, in terms of topological classifications, this black hole belongs to the class where RN black hole is situated \cite{wei2022black}. We also observed that with an increase in the size of $r_s\,(\rho_s)$, the zero point located at the smaller radius moves to a larger radius, while the zero point located at the larger radius moves to a smaller radius, reducing the distance between the two zero points.
\section{Conclusion}\label{Sec12}
In this study, we investigated a black hole surrounded by the  Dehnen-type dark matter halo.
Through our investigation of the black hole horizon, we identified the permissible parameter ranges that allow for the existence of black hole. Our thermodynamic study revealed that for specific selections of $r_s$ and $\rho_s$, a phase transition occurs at the Hawking temperature of the black hole. This phase transition prompted us to calculate the black hole remnant radius. Additionally, we analyzed light trajectories to determine the photon sphere and, subsequently, the shadow radius of the black hole. By comparing this value with Event Horizon Telescope (EHT) observations of Sgr A*, we established an upper limit for the parameters, confirming that the shadow remains within the permitted region, while no lower bound was found. We also derived the QNMs for both scalar and electromagnetic perturbations, observing that an increase in $r_s$ results in a decrease in both the real and imaginary parts of the QNMs. Notably, in electromagnetic scattering, both components exhibited larger values compared to scalar scattering. Furthermore, employing a topological approach, we examined the stability of the black hole's photon sphere, concluding that this black hole possesses only one unstable photon sphere, which contributes to the formation of its shadow. We investigated the phase transition at the black hole temperature under the condition $32\,\pi\,r_s^2\,\rho_s<1$, identifying a conventional critical point at maximum of temperature. Additionally, our study of the generalized free energy indicated two phase transitions; however, considering the total topological charge, the black hole aligns with the topological class of RN black holes.
These findings contribute significantly to our understanding of black hole physics and the interplay between dark matter and gravitational physics.
\hspace{4cm}
\section*{Acknowledgements}
The authors would like to thank the reviewer for their constructive comments. Additionally the authors express their gratitude to R. A. Konoplya, A. Zhidenko, and A. F. Zinhailo for the sharing their valuable \textit{Mathematica}\textregistered notebook with higher order WKB corrections \cite{Mathematica,konoplya2019higher}. F. S. and H. H. are grateful to Excellence project FoS UHK 2203/2025-2026 for the financial support. Furthermore, the research of H.H. was supported by the Q-CAYLE project, funded by the European Union-Next Generation UE/MICIU/Plan de Recuperacion, Transformacion y Resiliencia/Junta de Castilla y Leon (PRTRC17.11), and also by project PID2023-148409NB-I00, funded by MICIU/AEI/10.13039/501100011033. Financial support of the Department of Education of the Junta de Castilla y Leon and FEDER Funds is also gratefully acknowledged (Reference: CLU-2023-1-05).
\nocite{*}

\bibliographystyle{plain}

\begin{thebibliography}{99}
\section*{References}

\bibitem{bronnikov2013black}
Bronnikov, Kirill A., and Sergey G. Rubin. Black holes, cosmology and extra dimensions. 2013.

\bibitem{malafarina2017classical}
Malafarina, Daniele. "Classical collapse to black holes and quantum bounces: A review." Universe 3.2 (2017): 48.

\bibitem{volonteri2012black}
Volonteri, Marta, and Jillian Bellovary. "Black holes in the early Universe." Reports on Progress in Physics 75.12 (2012): 124901.

\bibitem{alexander2012drives}
Alexander, David M., and Ryan C. Hickox. "What drives the growth of black holes?." New Astronomy Reviews 56.4 (2012): 93-121.

\bibitem{volonteri2021origins}
Volonteri, Marta, Mélanie Habouzit, and Monica Colpi. "The origins of massive black holes." Nature Reviews Physics 3.11 (2021): 732-743.

\bibitem{rieger2011nonthermal}
Rieger, Frank M. "Nonthermal processes in black hole-jet magnetospheres—Invited review." International Journal of Modern Physics D 20.09 (2011): 1547-1596.

\bibitem{alexander2017stellar}
Alexander, Tal. "Stellar dynamics and stellar phenomena near a massive black hole." Annual Review of Astronomy and Astrophysics 55.1 (2017): 17-57.

\bibitem{blandford2019relativistic}
Blandford, Roger, David Meier, and Anthony Readhead. "Relativistic jets from active galactic nuclei." Annual Review of Astronomy and Astrophysics 57.1 (2019): 467-509.

\bibitem{frieman2008dark}
Frieman, Joshua A., Michael S. Turner, and Dragan Huterer. "Dark energy and the accelerating universe." Annu. Rev. Astron. Astrophys. 46.1 (2008): 385-432.

\bibitem{kauffmann1993formation}
Kauffmann, Guinevere, Simon DM White, and Bruno Guiderdoni. "The formation and evolution of galaxies within merging dark matter haloes." Monthly Notices of the Royal Astronomical Society 264.1 (1993): 201-218.

\bibitem{capozziello2012dark}
Capozziello, Salvatore, and Mariafelicia De Laurentis. "The dark matter problem from f (R) gravity viewpoint." Annalen der Physik 524.9‐10 (2012): 545-578.

\bibitem{bertone2018history}
Bertone, Gianfranco, and Dan Hooper. "History of dark matter." Reviews of Modern Physics 90.4 (2018): 045002.

\bibitem{capozziello2023modified}
Capozziello, Salvatore, et al. "Modified Kerr black holes surrounded by dark matter spike." arXiv e-prints (2023): arXiv-2311.

\bibitem{konoplya2022solutions}
Konoplya, R. A., and A. Zhidenko. "Solutions of the Einstein equations for a black hole surrounded by a galactic halo." The Astrophysical Journal 933.2 (2022): 166.

\bibitem{capozziello2023dark}
Capozziello, S., et al. "Dark matter spike around Bumblebee black holes." Journal of Cosmology and Astroparticle Physics 2023.05 (2023): 027.

\bibitem{shen2024analytical}
Shen, Zibo, et al. "Analytical models of supermassive black holes in galaxies surrounded by dark matter halos." Physics Letters B 855 (2024): 138797.

\bibitem{molla2025observable}
Molla, Niyaz Uddin, et al. "Observable signatures of RN black holes with dark matter halos via strong gravitational lensing and constraints from EHT observations." Physics of the Dark Universe 47 (2025): 101804.

\bibitem{abbasi2011search}
Abbasi, R., et al. "Search for Dark Matter from the Galactic Halo with the IceCube Neutrino Observatory." Phys. Rev. D 84.2 (2011).

\bibitem{ascasibar2004physical}
Ascasibar, Yago, et al. "On the physical origin of dark matter density profiles." Monthly Notices of the Royal Astronomical Society 352.4 (2004): 1109-1120.

\bibitem{merritt2006empirical}
Merritt, David, et al. "Empirical models for dark matter halos. I. Nonparametric construction of density profiles and comparison with parametric models." The Astronomical Journal 132.6 (2006): 2685.

\bibitem{catena2010novel}
Catena, Riccardo, and Piero Ullio. "A novel determination of the local dark matter density." Journal of Cosmology and Astroparticle Physics 2010.08 (2010): 004.

\bibitem{salucci2010dark}
Salucci, Paolo, et al. "The dark matter density at the Sun’s location." Astronomy \& Astrophysics 523 (2010): A83.


\bibitem{gohain2024thermodynamics}
Gohain, Mrinnoy M., Prabwal Phukon, and Kalyan Bhuyan. "Thermodynamics and null geodesics of a Schwarzschild black hole surrounded by a Dehnen type dark matter halo." Physics of the Dark Universe 46 (2024): 101683.

\bibitem{al2024quasinormal}
Al-Badawi, Ahmad, and Sanjar Shaymatov. "Quasinormal modes and shadow of Schwarzschild black holes embedded in a Dehnen-type dark matter halo exhibiting a cloud of strings." Communications in Theoretical Physics 77.3 (2024): 035402.

\bibitem{jha2025shadow}
Jha, Sohan Kumar. "Shadow, ISCO, quasinormal modes, Hawking spectrum, weak gravitational lensing, and parameter estimation of a Schwarzschild black hole surrounded by a Dehnen type dark matter halo." Journal of Cosmology and Astroparticle Physics 2025.03 (2025): 054.

\bibitem{dehnen1993family}
Dehnen, Walter. "A family of potential–density pairs for spherical galaxies and bulges." Monthly Notices of the Royal Astronomical Society 265.1 (1993): 250-256.

\bibitem{bronnikov2007regular}
Bronnikov, K. A., H. Dehnen, and V. N. Melnikov. "Regular black holes and black universes." General Relativity and Gravitation 39 (2007): 973-987.

\bibitem{khonji2024core}
Khonji, Nader, et al. "Core formation by binary scouring and gravitational wave recoil in massive elliptical galaxies." The Astrophysical Journal 974.2 (2024): 204.

\bibitem{alloqulov2025gravitational}
Alloqulov, Mirzabek, et al. "Gravitational waveforms from periodic orbits around a Schwarzschild black hole embedded in a Dehnen-type dark matter halo." arXiv preprint arXiv:2504.05236 (2025).

\bibitem{luo2025shadows}
Luo, Zuting, Meirong Tang, and Zhaoyi Xu. "Shadows and Observational Images of a Schwarzschild-like Black Hole Surrounded by a Dehnen-type Dark Matter Halo." arXiv preprint arXiv:2505.20115 (2025).

\bibitem{rani2025thermodynamic}
Rani, Shamaila, et al. "Thermodynamic and Shadow Analysis of Dehnen type Dark Matter Halo Corrected Schwarzschild Black Hole Surrounded by Thin Disk." Eur. Phys. J. C (2025) 85, 677.

\bibitem{konoplya2006stability}
Konoplya, R. A., and A. V. Zhidenko. "Stability and quasinormal modes of the massive scalar field around Kerr black holes." Physical Review D—Particles, Fields, Gravitation, and Cosmology 73.12 (2006): 124040.

\bibitem{nollert1999quantifying}
Nollert, Hans-Peter, and Richard H. Price. "Quantifying excitations of quasinormal mode systems." Journal of Mathematical Physics 40.2 (1999): 980-1010.

\bibitem{berti2003highly}
Berti, Emanuele, et al. "Highly damped quasinormal modes of Kerr black holes." Physical Review D 68.12 (2003): 124018.

\bibitem{ferrari1984new}
Ferrari, Valeria, and Bahram Mashhoon. "New approach to the quasinormal modes of a black hole." Physical Review D 30.2 (1984): 295.

\bibitem{yang2012quasinormal}
Yang, Huan, et al. "Quasinormal-mode spectrum of Kerr black holes and its geometric interpretation." Physical Review D—Particles, Fields, Gravitation, and Cosmology 86.10 (2012): 104006.

\bibitem{jaramillo2021pseudospectrum}
Jaramillo, José Luis, Rodrigo Panosso Macedo, and Lamis Al Sheikh. "Pseudospectrum and black hole quasinormal mode instability." Physical Review X 11.3 (2021): 031003.

\bibitem{konoplya2022nonoscillatory}
Konoplya, R. A., and A. Zhidenko. "Nonoscillatory gravitational quasinormal modes and telling tails for Schwarzschild–de Sitter black holes." Physical Review D 106.12 (2022): 124004.

\bibitem{heidari2024exploring}
Heidari, Narges, et al. "Exploring non-commutativity as a perturbation in the Schwarzschild black hole: quasinormal modes, scattering, and shadows." The European Physical Journal C 84.6 (2024): 566.

\bibitem{panotopoulos2019quasinormal}
Panotopoulos, Grigoris, and {\'A}ngel Rinc{\'o}n. "Quasinormal modes of regular black holes with non-linear electrodynamical sources." The European Physical Journal Plus 134.6 (2019): 300.

\bibitem{dutta2020revisiting}
Dutta Roy, Poulami, S. Aneesh, and Sayan Kar. "Revisiting a family of wormholes: geometry, matter, scalar quasinormal modes and echoes." The European Physical Journal C 80.9 (2020): 850.

\bibitem{lagos2020anomalous}
Lagos, Macarena, Pedro G. Ferreira, and Oliver J. Tattersall. "Anomalous decay rate of quasinormal modes." Physical Review D 101.8 (2020): 084018.


\bibitem{liang2025quasinormal}
Liang, Qi-Qi, Dong Liu, and Zheng-Wen Long. "Quasinormal Modes of Schwarzschild Black Holes in the Dehnen-(1, 4, 5/2) Type Dark Matter Halos." arXiv preprint arXiv:2505.15540 (2025).

\bibitem{al2025astrophysical}
Al-Badawi, Ahmad, and Sanjar Shaymatov. "Astrophysical properties of static black holes embedded in a Dehnen type dark matter halo with the presence of quintessential field." arXiv preprint arXiv:2501.15397 (2025).

\bibitem{hamil2025geodesics}
Hamil, B., Ahmad Al-Badawi, and B. C. L{\"u}tf{\"u}o{\u{g}}lu. "Geodesics and scalar perturbations of Schwarzschild black holes embedded in a Dehnen-type dark matter halo with quintessence." arXiv preprint arXiv:2505.18611 (2025).

\bibitem{uktamov2025static}
Uktamov, Uktamjon, Sanjar Shaymatov, and Bobomurat Ahmedov. "Static Black Hole Solution with a Dark Matter Halo." arXiv preprint arXiv:2505.20031 (2025).

\bibitem{navarro1996structure}
Navarro, Julio F. "The structure of cold dark matter halos." Symposium-international astronomical union. Vol. 171. Cambridge University Press, 1996.

\bibitem{moore1999cold}
Moore, Ben, et al. "Cold collapse and the core catastrophe." Monthly Notices of the Royal Astronomical Society 310.4 (1999): 1147-1152.

\bibitem{kravtsov1998cores}
Kravtsov, Andrey V., et al. "The cores of dark matter-dominated galaxies: theory versus observations." The Astrophysical Journal 502.1 (1998): 48.

\bibitem{mo2010galaxy}
Mo, Houjun, Frank Van den Bosch, and Simon White. Galaxy formation and evolution. Cambridge University Press, 2010.

\bibitem{al2024schwarzschild}
Al-Badawi, Ahmad, Sanjar Shaymatov, and Yassine Sekhmani. "Schwarzschild black hole in galaxies surrounded by a dark matter halo." Journal of Cosmology and Astroparticle Physics 2025.02 (2025): 014.

\bibitem{hawking1974black}
Hawking, Stephen W. "Black hole explosions?." Nature 248.5443 (1974): 30-31.

\bibitem{delhom2019absorption}
Delhom, Adria, et al. "Absorption by black hole remnants in metric-affine gravity." Physical Review D 100.2 (2019): 024016.

\bibitem{ong2024case}
Ong, Yen Chin. "The Case For Black Hole Remnants: A Review." arXiv preprint arXiv:2412.00322 (2024).

\bibitem{araujo2023thermodynamics}
Ara{\'u}jo Filho, A. A., et al. "Thermodynamics and evaporation of a modified Schwarzschild black hole in a non–commutative gauge theory." Physics Letters B 838 (2023): 137744.

\bibitem{davies1978thermodynamics}
Davies, Paul CW. "Thermodynamics of black holes." Reports on Progress in Physics 41.8 (1978): 1313.

\bibitem{tsallis2019black}
Tsallis, Constantino. "Black hole entropy: A closer look." Entropy 22.1 (2019): 17.

\bibitem{thomas2012thermodynamics}
Thomas, Bouetou Bouetou, Mahamat Saleh, and Timoleon Crepin Kofane. "Thermodynamics and phase transition of the Reissner–Nordstr{\"o}m black hole surrounded by quintessence." General Relativity and Gravitation 44 (2012): 2181-2189.

\bibitem{de2012black}
de la Cruz-Dombriz, Alvaro, and Diego S{\'a}ez-G{\'o}mez. "Black holes, cosmological solutions, future singularities, and their thermodynamical properties in modified gravity theories." Entropy 14.9 (2012): 1717-1770.

\bibitem{compere2007symmetries}
Compere, Geoffrey. "Symmetries and conservation laws in Lagrangian gauge theories with applications to the mechanics of black holes and to gravity in three dimensions." arXiv preprint arXiv:0708.3153 (2007).

\bibitem{frolov2003particle}
Frolov, Valeri, and Dejan Stojkovi{\'c}. "Particle and light motion in a space-time of a five-dimensional rotating black hole." Physical Review D 68.6 (2003): 064011.

\bibitem{he2016gravitational}
He, Guansheng, and Wenbin Lin. "Gravitational deflection of light and massive particles by a moving Kerr–Newman black hole." Classical and Quantum Gravity 33.9 (2016): 095007.

\bibitem{gralla2020shape}
Gralla, Samuel E., Alexandru Lupsasca, and Daniel P. Marrone. "The shape of the black hole photon ring: A precise test of strong-field general relativity." Physical Review D 102.12 (2020): 124004.

\bibitem{perlick2022calculating}
Perlick, Volker, and Oleg Yu Tsupko. "Calculating black hole shadows: Review of analytical studies." Physics Reports 947 (2022): 1-39.

\bibitem{virbhadra2000schwarzschild}
Virbhadra, Kumar Shwetketu, and George FR Ellis. "Schwarzschild black hole lensing." Physical Review D 62.8 (2000): 084003.

\bibitem{claudel2001geometry}
Claudel, Clarissa-Marie, Kumar Shwetketu Virbhadra, and George FR Ellis. "The geometry of photon surfaces." Journal of Mathematical Physics 42.2 (2001): 818-838.

\bibitem{zakharov2012shadows}
Zakharov, Alexander F., et al. "Shadows as a tool to evaluate black hole parameters and a dimension of spacetime." New Astronomy Reviews 56.2-3 (2012): 64-73.

\bibitem{atamurotov2015optical}
Atamurotov, Farruh, Bobomurat Ahmedov, and Ahmadjon Abdujabbarov. "Optical properties of black holes in the presence of a plasma: The shadow." Physical Review D 92.8 (2015): 084005.

\bibitem{jusufi2020quasinormal}
Jusufi, Kimet. "Quasinormal modes of black holes surrounded by dark matter and their connection with the shadow radius." Physical Review D 101.8 (2020): 084055.

\bibitem{perlick2018black}
Perlick, Volker, Oleg Yu Tsupko, and Gennady S. Bisnovatyi-Kogan. "Black hole shadow in an expanding universe with a cosmological constant." Physical Review D 97.10 (2018): 104062.

\bibitem{vagnozzi2023horizon}
S. Vagnozzi, R. Roy, Y.-D. Tsai, L. Visinelli, M. Afrin,
A. Allahyari, P. Bambhaniya, D. Dey, S. G. Ghosh, P. S.
Joshi, et al., “Horizon-scale tests of gravity theories and
fundamental physics from the event horizon telescope image
of sagittarius a,” Classical and Quantum Gravity,
vol. 40, no. 16, p. 165007, 2023.

\bibitem{nollert1999quasinormal}
Nollert, Hans-Peter. "Quasinormal modes: the characteristicsound'of black holes and neutron stars." Classical and Quantum Gravity 16.12 (1999): R159.

\bibitem{cavalcanti2022echoes}
Cavalcanti, R. T., R. C. de Paiva, and R. da Rocha. "Echoes of the gravitational decoupling: scalar perturbations and quasinormal modes of hairy black holes." The European Physical Journal Plus 137.10 (2022): 1185.

\bibitem{alpeggiani2017quasinormal}
Alpeggiani, Filippo, et al. "Quasinormal-mode expansion of the scattering matrix." Physical Review X 7.2 (2017): 021035.

\bibitem{lalanne2019quasinormal}
Lalanne, Philippe, et al. "Quasinormal mode solvers for resonators with dispersive materials." JOSA A 36.4 (2019): 686-704.

\bibitem{sauvan2022normalization}
Sauvan, Christophe, et al. "Normalization, orthogonality, and completeness of quasinormal modes of open systems: the case of electromagnetism." Optics Express 30.5 (2022): 6846-6885.

\bibitem{vishveshwara1970scattering}
Vishveshwara, C. V. "Scattering of gravitational radiation by a Schwarzschild black-hole." Nature 227.5261 (1970): 936-938.

\bibitem{kokkotas1999quasi}
Kokkotas, Kostas D., and Bernd G. Schmidt. "Quasi-normal modes of stars and black holes." Living Reviews in Relativity 2 (1999): 1-72.

\bibitem{hegde2019quasinormal}
Hegde, Suraj S., et al. "Quasinormal modes and the Hawking-Unruh effect in quantum Hall systems: lessons from black hole phenomena." Physical review letters 123.15 (2019): 156802.

\bibitem{blazquez2019quasinormal}
Bl{\'a}zquez-Salcedo, Jose Luis, Sarah Kahlen, and Jutta Kunz. "Quasinormal modes of dilatonic Reissner–Nordstr{\"o}m black holes." The European Physical Journal C 79 (2019): 1-15.

\bibitem{cai2021quasinormal}
Cai, Xin-Chang, and Yan-Gang Miao. "Quasinormal modes and shadows of a new family of Ay{\'o}n-Beato-Garc{\'\i}a black holes." Physical Review D 103.12 (2021): 124050.

\bibitem{franchini2024testing}
Franchini, Nicola, and Sebastian H. Völkel. "Testing general relativity with black hole quasi-normal modes." Recent Progress on Gravity Tests: Challenges and Future Perspectives. Singapore: Springer Nature Singapore, 2024. 361-416.

\bibitem{heidari2023investigation}
Heidari, N., and H. Hassanabadi. "Investigation of the quasinormal modes of a Schwarzschild black hole by a new generalized approach." Physics Letters B 839 (2023): 137814.

\bibitem{kovtun2005quasinormal}
Kovtun, Pavel K., and Andrei O. Starinets. "Quasinormal modes and holography." Physical Review D—Particles, Fields, Gravitation, and Cosmology 72.8 (2005): 086009.

\bibitem{konoplya2011quasinormal}
Konoplya, R. A., and Alexander Zhidenko. "Quasinormal modes of black holes: From astrophysics to string theory." Reviews of Modern Physics 83.3 (2011): 793-836.

\bibitem{volkel2022quasinormal}
V{\"o}lkel, Sebastian H. "Quasinormal modes from bound states: The numerical approach." Physical Review D 106.12 (2022): 124009.

\bibitem{cho2012new}
Cho, H. T., et al. "A new approach to black hole quasinormal modes: a review of the asymptotic iteration method." Advances in Mathematical Physics 2012.1 (2012): 281705.

\bibitem{schutz1985black}
Schutz, Bernard F., and Clifford M. Will. "Black hole normal modes: a semianalytic approach." The Astrophysical Journal 291 (1985): L33-L36.

\bibitem{guinn1990high}
Guinn, James Williams. High-overtone black-hole normal modes: A WKB contour-integral approach. Washington University in St. Louis, 1990.

\bibitem{konoplya2021black}
Konoplya, R. A. "Black holes in galactic centers: quasinormal ringing, grey-body factors and Unruh temperature." Physics Letters B 823 (2021): 136734.


\bibitem{leaver1986spectral}
Leaver, Edward W. "Spectral decomposition of the perturbation response of the Schwarzschild geometry." Physical Review D 34.2 (1986): 384.

\bibitem{leaver1990quasinormal}
Leaver, Edward W. "Quasinormal modes of Reissner-Nordström black holes." Physical Review D 41.10 (1990): 2986.
%
\bibitem{konoplya2003quasinormal}
Konoplya, R. A. "Quasinormal behavior of the D-dimensional Schwarzschild black hole and the higher order WKB approach." Physical Review D 68.2 (2003): 024018.


\bibitem{matyjasek2017quasinormal}
Matyjasek, Jerzy, and Michał Opala. "Quasinormal modes of black holes: The improved semianalytic approach." Physical Review D 96.2 (2017): 024011.
%
\bibitem{konoplya2019higher}
Konoplya, R. A., A. Zhidenko, and A. F. Zinhailo. "Higher order WKB formula for quasinormal modes and grey-body factors: recipes for quick and accurate calculations." Classical and Quantum Gravity 36.15 (2019): 155002.

\bibitem{Mathematica}
\href{https://goo.gl/nykYGL}{
The open access \textit{Mathematica}\textregistered package with the WKB formula of
13th order and Padé approximations.}

\bibitem{du2024topological}
Du, Yongbin, Haida Li, and Xiangdong Zhang. "Topological classes of BTZ black holes." Symmetry 16.12 (2024): 1577.

\bibitem{s2024effective}
S. Afshar, Mohammad Ali, and Jafar Sadeghi. "Effective potential and topological photon spheres: a novel approach to black hole parameter classification." Chinese Physics C (2024).

\bibitem{wu2024novel}
Wu, Di, et al. "Novel Topological Classes in Black Hole Thermodynamics." arXiv preprint arXiv:2411.10102 (2024).

\bibitem{hazarika2024thermodynamic}
Hazarika, Bidyut, B. Eslam Panah, and Prabwal Phukon. "Thermodynamic topology of topological charged dilatonic black holes." The European Physical Journal C 84.11 (2024): 1-18.

\bibitem{malik2024thermodynamic}
Malik, Adnan, Aqsa Mehmood, and M. Umair Shahzad. "Thermodynamic topological classification of higher dimensional and massive gravity black holes." Annals of Physics 463 (2024): 169617.

\bibitem{barzi2024renyi}
Barzi, F., H. El Moumni, and K. Masmar. "R{\'e}nyi topology of charged-flat black hole: Hawking-Page and Van-der-Waals phase transitions." Journal of High Energy Astrophysics 42 (2024): 63-86.

\bibitem{liu2024light}
Liu, Wentao, Di Wu, and Jieci Wang. "Light rings and shadows of static black holes in effective quantum gravity." Physics Letters B 858 (2024): 139052.

\bibitem{sadeghi2024thermodynamic}
Sadeghi, Jafar, et al. "Thermodynamic topology and photon spheres in the hyperscaling violating black holes." Astroparticle Physics 156 (2024): 102920.

\bibitem{Umair}
Shahzad, M. Umair, et al. "Topological Photon Spheres of Regular Non-Minimal Magnetic Black Holes." International Journal of Theoretical Physics 64.6 (2025): 1-13.

\bibitem{wei2020topological}
Wei, Shao-Wen. "Topological charge and black hole photon spheres." Physical Review D 102.6 (2020): 064039.

\bibitem{wu2023classifying}
Wu, Di. "Classifying topology of consistent thermodynamics of the four-dimensional neutral Lorentzian NUT-charged spacetimes." The European Physical Journal C 83.5 (2023): 365.

\bibitem{sadeghi2024role}
Sadeghi, Jafar, and Mohammad Ali S. Afshar. "The role of topological photon spheres in constraining the parameters of black holes." Astroparticle Physics (2024): 102994.

\bibitem{zhu2024topological}
Zhu, Xiao-Dan, Di Wu, and Dan Wen. "Topological classes of thermodynamics of the rotating charged AdS black holes in gauged supergravities." Physics Letters B 856 (2024): 138919.

\bibitem{rizwan2025universal}
Rizwan, Muhammad, Mubasher Jamil, and M. Z. A. Moughal. "Universal thermodynamic topological classes of black holes in perfect fluid dark matter background." arXiv preprint arXiv:2501.04739 (2025).

\bibitem{chen2024thermal}
Chen, H., et al. "Thermal, topological, and scattering effects of an AdS charged black hole with an antisymmetric tensor background." arXiv preprint arXiv:2408.03090 (2024).

\bibitem{yerra2022topology}
Yerra, Pavan Kumar, Chandrasekhar Bhamidipati, and Sudipta Mukherji. "Topology of critical points and Hawking-Page transition." Physical Review D 106.6 (2022): 064059.

\bibitem{wei2022topology}
Wei, Shao-Wen, and Yu-Xiao Liu. "Topology of black hole thermodynamics." Physical Review D 105.10 (2022): 104003.

\bibitem{chen2024thermodynamic}
Chen, Hao, et al. "Thermodynamic topology of phantom AdS black holes in massive gravity." Physics of the Dark Universe 46 (2024): 101617.

\bibitem{dong2025thermodynamic}
Dong, Shi-Hai, et al. "Thermodynamic properties and topological charge of a static black hole in loop quantum gravity." Physics of the Dark Universe (2025): 101962.

\bibitem{yerra2023topology}
Yerra, Pavan Kumar, Chandrasekhar Bhamidipati, and Sudipta Mukherji. "Topology of Hawking-Page transition in Born-Infeld AdS black holes." Journal of Physics: Conference Series. Vol. 2667. No. 1. IOP Publishing, 2023.

\bibitem{wu2023topologicall}
Wu, Di, and Shuang-Qing Wu. "Topological classes of thermodynamics of rotating AdS black holes." Physical Review D 107.8 (2023): 084002.

\bibitem{wei2024universal}
Wei, Shao-Wen, Yu-Xiao Liu, and Robert B. Mann. "Universal topological classifications of black hole thermodynamics." Physical Review D 110.8 (2024): L081501.

\bibitem{wei2022black}
Wei, Shao-Wen, Yu-Xiao Liu, and Robert B. Mann. "Black hole solutions as topological thermodynamic defects." Physical Review Letters 129.19 (2022): 191101.

\bibitem{wu2023topological}
Wu, Di. "Topological classes of thermodynamics of the four-dimensional static accelerating black holes." Physical Review D 108.8 (2023): 084041.

\bibitem{liu2023topological}
Liu, Conghua, and Jin Wang. "Topological natures of the Gauss-Bonnet black hole in AdS space." Physical Review D 107.6 (2023): 064023.

\bibitem{yasir2024topological}
Yasir, Muhammad, Xia Tiecheng, and Abdul Jawad. "Topological charges via Barrow entropy of black hole in metric-affine gravity." The European Physical Journal C 84.9 (2024): 1-15.

\bibitem{fan2023topological}
Fan, Zhong-Ying. "Topological interpretation for phase transitions of black holes." Physical Review D 107.4 (2023): 044026.%

\bibitem{wu2024topological}
Wu, Di, et al. "Topological classes of thermodynamics of the static multi-charge AdS black holes in gauged supergravities: novel temperature-dependent thermodynamic topological phase transition." Journal of High Energy Physics 2024.6 (2024): 1-35.

\end{thebibliography}

\end{document}